# Incommensurate antiferromagnetism in UTe$_2$ under pressure

W. Knafo,[1,*] T. Thebault[1], P. Manuel[2], D.D. Khalyavin[2], F. Orlandi[2], E. Ressouche[3], K. Beauvois[3], G. Lapertot[4], K. Kaneko[5,6], D. Aoki[7], D. Braithwaite[4], G. Knebel[4], S. Raymond[3]

[1] *Laboratoire National des Champs Magnétiques Intenses - EMFL, CNRS, Université Grenoble Alpes, INSA-T, Université Toulouse 3, 31400 Toulouse, France.*
[2] *ISIS Pulsed Neutron and Muon Source, STFC Rutherford Appleton Laboratory, Harwell Science and Innovation Campus, Oxfordshire OX11 0QX, United Kingdom.*
[3] *Université Grenoble Alpes, CEA, IRIG, MEM, MDN, 38000 Grenoble, France.*
[4] *Université Grenoble Alpes, CEA, Grenoble INP, IRIG, PHELIQS, 38000 Grenoble, France.*
[5] *Materials Sciences Research Center, Japan Atomic Energy Agency, Tokai, Ibaraki 319-1195, Japan*
[6] *Advanced Science Research Center, Japan Atomic Energy Agency, Tokai, Ibaraki 319-1195, Japan*
[7] *Institute for Materials Research, Tohoku University, Ibaraki 311-1313, Japan*

*\* Corresponding author: william.knafo@lncmi.cnrs.fr*

**Abstract**

The discovery of multiple superconducting phases in UTe$_2$ boosted research on correlated-electron physics. This heavy-fermion paramagnet was rapidly identified as a reference compound to study the interplay between magnetism and unconventional superconductivity with multiple degrees of freedom. The proximity to a ferromagnetic quantum phase transition was initially proposed as a driving force to triplet-pairing superconductivity. However, we find here that long-range incommensurate antiferromagnetic order is established under pressure. The propagation vector $\mathbf{k_m} = (0.07, 0.33, 1)$ of the antiferromagnetic phase is close to a wavevector where antiferromagnetic fluctuations have previously been observed at ambient pressure. These elements support that UTe$_2$ is a nearly-antiferromagnet at ambient pressure. Our work appeals for theories modelling the evolution of the magnetic interactions and electronic properties, driving a correlated paramagnetic regime at ambient pressure to a long-range antiferromagnetic order under pressure. A deeper understanding of itinerant-$f$-electrons magnetism in UTe$_2$ will be a key for describing its unconventional superconducting phases.

**Introduction**

UTe$_2$ offers a unique chance to investigate the interplay between magnetism and unconventional superconductivity in correlated-electron physics [1,2]. This heavy-fermion material crystallizes in the orthorhombic structure Immm and is paramagnetic down to the lowest temperatures [3]. Its electronic and magnetic properties are anisotropic, and rich three-dimensional ($p,H,T$) phase diagrams, where $p$ is the pressure, $H$ is the magnetic field, and $T$ is the temperature, have been determined for different directions of the magnetic field [4,5,6,7,8,9,10,11,12,13]. An unprecedented number of superconducting phases were found in the vicinity of magnetic quantum phase transitions, i.e., magnetic phase transitions in the limit $T \rightarrow 0$, induced under pressure or magnetic field. We count so far at least three superconducting phases which can be tuned under pressure at zero field [4,5,6,14], but also several superconducting phases stabilized under magnetic fields **H** ∥ **a** [10], **b** [7,8,15,16], **c** [11,12], and **H** tilted by ≈ 30° from **b** towards **c** [8,9,13], at ambient pressure or under pressure. An unconventional nature was highlighted and a magnetically-mediated mechanism is suspected for most of these superconducting phases. Large upper superconducting critical fields (far above the Pauli limitation expected for a weak-coupling limit with a Landé factor $g = 2$), but also the stabilization of new superconducting phases under a magnetic field, were identified as supports for triplet superconductivity [1,4,5]. A triplet nature of the superconducting phases was also proposed from nuclear-magnetic-resonance (NMR) Knight-shift measurements [17,18]. The interest for UTe$_2$ was reinforced by the proposition of a chiral nature of superconductivity [19] and the identification of a charge-density wave (or pair-density wave) [20,21] from scanning-tunnelling-microscopy (STM) experiments.

A remaining challenge is to determine the relation between the magnetic correlations and superconductivity in different pressure and magnetic-field conditions. UTe$_2$ was initially presented as a nearly-ferromagnet and ferromagnetic fluctuations were proposed to drive an unconventional triplet mechanism for the superconducting phase SC1, which develops at ambient pressure and zero magnetic field [4] [see Figure 1(a)]. However, no trace of ferromagnetic fluctuations or long-range ferromagnetic order has been observed so far. An initial claim for a signature of ferromagnetic fluctuations from muon-spin-relaxation measurements [22] was attributed to a disorder effect in a further study by the same group [23]. On the contrary, low-dimensional antiferromagnetic fluctuations were identified by inelastic neutron scattering [24,25,26], indicating a possible nearby quantum antiferromagnetic instability. Although the observed fluctuations are antiferromagnetic, a ferromagnetic coupling between the two closest U atoms, which constitute the rungs of U two-legs magnetic ladders in the structure of UTe$_2$ [see Figure 1(b)], was deduced in [25]. The ladder structure and the fact that U atoms do not lie on inversion-symmetry centers were proposed to play a role for a triplet superconducting pairing in [27,28]. A gapping of the antiferromagnetic fluctuations was also evidenced in the phase SC1 [29,30], demonstrating their intimate relation with the superconducting mechanism.

Figure 1(a) presents the pressure – temperature phase diagram of UTe$_2$ at zero magnetic field constructed using data from Refs. [6,11,31,32]. A quantum phase transition occurs at the critical pressure $p_c$ ≈ 1.5-1.7 GPa. At ambient pressure, a correlated paramagnetic (CPM) regime is characterized by two temperatures scales $T_\chi^{max}$ = 35 K and $T^*$ = 15 K [33,34], which decrease continuously with $p$ up to $p_c$ [11,32,35]. A CPM regime can be identified in many heavy-fermion paramagnets, where a broad maximum, at the temperature $T_\chi^{max}$, in the magnetic susceptibility indicates the onset of electronic correlations [36]. A signature of these correlations are the antiferromagnetic fluctuations, which were observed by inelastic neutron scattering in UTe$_2$, at the incommensurate wavevector $\mathbf{k}_1 = (0\ 0.57\ 0)$, [24,25,26] (see also NMR experiments [37]) and in other heavy-fermion paramagnets [36]. For $p < p_c$, superconductivity is reinforced with pressure, as indicated by the increase of the superconducting temperature $T_{sc}$ from 1.5-2.1 K (depending on the sample quality [38,39]) at ambient pressure to $\approx$ 3 K near $p_c$ [6]. The increase of $T_{sc}$ is associated with the emergence of a second superconducting phase SC2 under pressure. SC2 suddenly disappears at pressures $p > p_c$, where a magnetically-ordered phase is stabilized below a critical temperature reaching $\approx$ 3.5 - 4 K near $p_c$, and which increases with $p$. The microscopic nature of the magnetically-ordered phase has not been determined and different propositions were made from bulk electrical-transport and thermodynamics measurements. Following the observation of a hysteresis in electrical-resistivity data, the magnetically-ordered phase was proposed to be ferromagnetic in [4,15,40]. On the contrary, this phase was proposed to be antiferromagnetic from other features observed at the transition temperature in the electrical resistivity [6,41] and in the magnetic susceptibility [31]. In addition, the phase boundaries of the magnetically-ordered phase at a critical field $H_c$ for all magnetic field directions, but also intermediate field-induced transitions at magnetic fields $H_{ri} < H_c$ were taken as the indication for an antiferromagnetic structure [11,12,32]. For $p > p_c$, a higher-temperature scale $T_{WMO}$, of $\approx$ 10-15 K near $p_c$, was identified from broad anomalies in magnetic-susceptibility, heat-capacity and electrical-resistivity measurements [11,12,31,41]. This regime was proposed to result from a weak magnetic order (WMO), i.e., short-range magnetic correlations without long-range magnetic order. Equivalently, it can also be labelled as a CPM regime, since its upper temperature corresponds to a crossover at which the magnetic susceptibilities $\chi_a$ and $\chi_c$, measured in magnetic fields $\mathbf{H} \parallel \mathbf{a}$ and $\mathbf{c}$, respectively, present a maximum. This high-pressure and high-temperature correlated regime is noted WMO/CPM here.

In this Letter, we present a neutron-diffraction study of UTe$_2$ in its pressure-induced magnetically-ordered phase. Experiments were performed under a pressure $p = 1.8$ GPa and in zero magnetic field. Magnetic Bragg peaks are observed at temperatures below $T_N \approx 3.5$ K. They are the signature of an antiferromagnetic (AF) order with the incommensurate wavevector $\mathbf{k_m} = (0.07, 0.33, 1)$ and are associated with a magnetic moment $\mu_m \geq 0.3 \pm 0.05\ \mu_B$/U. Our work shows that superconductivity in UTe$_2$ develops in the vicinity of a long-range antiferromagnetic phase, which differs from the initial proposition of a nearby ferromagnetic phase [4].

**Results**

Figure 2 presents the signatures of magnetic Bragg peaks measured at the momentum transfers $\mathbf{Q} = (\pm 0.07, 0.67, 0)$. They were extracted using the Laue time-of-flight spectrometer WISH (spallation source ISIS, Didcot) in a first experimental configuration with the basal scattering plane perpendicular to the axis $\mathbf{a}$ of the crystal and an angle of 14.5 ° between the incident neutron beam and the axis $\mathbf{c}$ of the crystal, at the temperature $T = 1.5$ K. Neutron-scattered intensity maps are shown in the $(Q_h, Q_k, 0)$ and $(0.07, Q_k, Q_l)$ planes in the Panels (a-b), respectively, and $Q_k$, $Q_h$, and $Q_l$ scans extracted from these maps are shown in the Panels (c-e), respectively. Bragg peaks are fitted by a Gaussian function. Figure S2 in the Supplementary Information shows that, in a second experimental configuration (angle of 38.5 ° between the incident neutron beam and $\mathbf{c}$), signatures of Bragg peaks were also observed at the momentum transfers $\mathbf{Q} = (\pm 0.07, 1.33, 0)$ at the temperature $T = 1.5$ K. We note that, due to the limited number of observed out-of-plane structural reflections, a small error in the orientation matrix used here is responsible for the slightly negative values of $Q_l$ at the maxima in the $Q_l$ scans [see Figures 2(e) and S2(e)]. The absence of Bragg peak with a positive value of $Q_l$ in these scans indicates that the peak expected for a perfect orientation matrix should be centered around $Q_l = 0$. The formula $\mathbf{Q} = \boldsymbol{\tau} \pm \mathbf{k}$ relates a momentum transfer **Q** to pairs of wavevectors ±**k** defined within the first Brillouin zone (FBZ) and the position **τ** of a structural Bragg peak. The four observed Bragg peaks correspond to wavevectors equivalent to $\mathbf{k_m} = (0.07, 0.33, 1)$, which lies on the border of the FBZ [see Figure 4(d)]. The observation of magnetic Bragg peaks at momentum transfers of components $Q_l = 0$ further indicates that the two closest U atoms, which form the rungs of the ladders, have in-phase magnetic moments (local ferromagnetic arrangement), presumably due to a ferromagnetic coupling between them. This information can be extracted since these two U atoms, distant along the direction **c**, belong to the same primitive cell (see [25] and Supplementary Information).

Figures 3 (a-c) present $Q_h$, $Q_k$, and $Q_l$ scans, respectively, extracted in the vicinity of the momentum transfer $\mathbf{Q} = (0.07, 0.67, 0)$, for different temperatures ranging from 1.5 K to 15 K. The Bragg peak disappears at temperatures higher than 3.5 K, as expected for a magnetic signal. This temperature scale coincides with the transition temperature of the pressure-induced phase reported in bulk heat-capacity and electrical-resistivity measurements [6]. This temperature is, thus, the signature of the antiferromagnetic ordering identified here and is labelled as the Néel temperature $T_N \approx 3.5$ K. Interestingly, the position of the Bragg peak slightly changes with temperature. The components $Q_h$ and $Q_k$ of the momentum transfer **Q**, at which the scattered intensity is maximal, and therefore the components $k_{mh}$ and $k_{mk}$ of the magnetic wavevector $\mathbf{k_m}$, are slightly affected by the temperature [see Figures 4(b,c)]. The first component of $\mathbf{k_m}$ varies from $k_{mh} = 0.069$ at $T = 1.5$ K to $k_{mh} = 0.076$ at $T = 3$ K while the second component of $\mathbf{k_m}$ varies from $k_{mk} = 0.333$ at $T = 1.5$ K to $k_{mk} = 0.335$ at $T = 3$ K. Within the experimental resolution, we cannot see a variation with temperature of the third component $k_{ml}$ of $\mathbf{k_m}$.

The temperature dependence of the intensity $I_m$ of the magnetic Bragg peak at $\mathbf{Q} = (0.07, 0.67, 0)$ is presented in Figure 4(a). $I_m$ is enhanced below the temperature $T_N \approx 3.5$ K. It is proportional to the square of the component $\mu_m^\perp$ perpendicular to **Q** of the moments $\boldsymbol{\mu_m}$ ordered with the wavevector $\mathbf{k_m}$. The moment amplitude $\mu_m$ is therefore larger or equal to $\mu_m^\perp$. At $T = 1.6$ K, a comparison of the intensities of the magnetic Bragg peaks measured at $\mathbf{Q} = (0.07, 0.67, 0)$ and $\mathbf{Q} = (0.07, 1.33, 0)$ with the intensities of a series of eight structural Bragg peaks, measured complementarily using the two-axes neutron diffractometer D23 (reactor ILL, Grenoble), permits to extract the amplitude of the moment component $\mu_m^\perp = 0.3 \pm 0.05$ $\mu_B$/U perpendicular to **Q**. $\mu_m^\perp$ almost corresponds to the projection of the ordered magnetic moment $\boldsymbol{\mu_m}$ in the plane perpendicular to **b**.

**Discussion**

*Magnetic structure.* The collection of magnetic Bragg peaks observed here is not sufficient to completely solve the magnetic structure of UTe$_2$ under pressure. The direction of the magnetic moments cannot be extracted, and we cannot determine whether they order colinearly within a spin-density-wave modulation or non-colinearly within a helical modulation (see Supplementary Information). In addition, the question whether the antiferromagnetic phase leads to the formation of several single-**k** domains or a single multi-**k** domain cannot be answered. Two groups of wavevectors are identified: a first group of wavevectors equivalent to $\mathbf{k_{m11}} = (0.07, 0.33, 1)$ and a second group of wavevectors equivalent to $\mathbf{k_{m21}} = (-0.07, 0.33, 1)$ [see Figure 4(d)]. Each group of wavevectors can drive to the formation of a single-**k** domain, ending in two single-**k** domains in the crystal. Alternatively, a unique multi-**k** domain consisting of a superposition of contributions at all wavevectors could also be present. To solve the magnetic structure of UTe$_2$ in its pressure-induced antiferromagnetic phase, further neutron diffraction experiments at other magnetic Bragg reflections, in particular at larger $Q = |\mathbf{Q}|$ values, are needed. We note that the work presented here was already an experimental tour de force, due to the strong neutron absorption and geometrical constraints of the cell used to reach high pressures, combined with the small amplitude of the measured ordered magnetic moment. The full determination of the magnetic structure of UTe$_2$ in its high-pressure antiferromagnetic phase will constitute an experimental challenge to overcome in the next years. Interestingly, we have observed a small variation of the antiferromagnetic wavevector when the temperature is varied. While the first component $k_{mh}$ of $\mathbf{k_m}$ remains incommensurate at the different temperatures probed here, the second component $k_{mk}$ of $\mathbf{k_m}$ is close to the commensurate value 1/3 at $T = 1.5$ K and a small deviation from this value is observed when the temperature is increased. A similar change with temperature of an incommensurate magnetic wavevector was already observed in other antiferromagnets, see for instance CeCu$_2$Ge$_2$ [42]. The transitions reported at magnetic fields $H_{ri}$ in magnetic fields applied along $\mathbf{b}$ (with or without tilt) and $\mathbf{c}$ [11,12,32] are presumably related to subtle changes of the antiferromagnetic structure, such as a moment reorientation, a change of magnetic wavevector, and/or a domain selection. In various heavy-fermion antiferromagnets with an incommensurate magnetic structure, a change of the magnetic wavevector was found to be induced by a magnetic field (see again CeCu$_2$Ge$_2$ [43]). A further target could be to study how the pressure-induced magnetic structure of UTe$_2$ is modified in a magnetic field.

*Magnetic fluctuations.* Within a conventional description of quantum magnetic criticality, antiferromagnetic fluctuations in the paramagnetic state can be a precursor of an antiferromagnetic order [44,45]. At low temperature, their intensity grows and their characteristic energy scale decreases when the magnetic quantum phase transition is approached, ending in their transformation into a long-range magnetic order beyond the phase transition. They can be probed directly by inelastic neutron scattering and NMR experiments, or indirectly by extracting the low-temperature Fermi-liquid coefficients $\gamma$ from heat-capacity measurements and $A$ from electrical-resistivity measurements. Critical antiferromagnetic-order-parameter fluctuations were identified by inelastic neutron scattering at a magnetic quantum phase transition in the heavy fermion antiferromagnetic system Ce$_{1-x}$La$_x$Ru$_2$Si$_2$ [46] (see also CeCo(In$_{1-x}$Hg$_x$)$_5$ [47]). In UTe$_2$, antiferromagnetic fluctuations with the wavevector $\mathbf{k_1} = (0, 0.57, 0)$ were observed at ambient pressure [24,25,26]. A maximum of the electrical-resistivity coefficient $A$ was also reported at the critical pressure [6], indicating that critical magnetic fluctuations precede the onset of antiferromagnetic order with wavevector $\mathbf{k_m} = (0.07, 0.33, 1)$. These phenomena support that UTe$_2$ is a nearly-antiferromagnet in its CPM regime at ambient pressure. Figure 4(e) further shows that $\mathbf{k_1}$ and its equivalent positions in the reciprocal space are close to $\mathbf{k_m}$ and its equivalent positions in the reciprocal space [for instance, $\mathbf{k_m}$ is near to $\mathbf{k_1'} = (0, 0.43, 1) = \boldsymbol{\tau} - \mathbf{k_1}$, which is equivalent to $-\mathbf{k_1}$ and can be obtained via a reciprocal translation with vector $\boldsymbol{\tau} = (0,1,1)$]. The closeness of $\mathbf{k_1}$ and $\mathbf{k_m}$ suggests that the antiferromagnetic fluctuations with wavevector $\mathbf{k_1}$ may be a precursor of the antiferromagnetic long-range order with wavevector $\mathbf{k_m}$. In addition to the increase of the strength of the antiferromagnetic fluctuations, we speculate that a modification, from $\mathbf{k_1}$ to $\mathbf{k_m}$, of their associated wavevector may be induced under pressure, ending in the stabilization of long-range magnetic order with $\mathbf{k_m}$ for $p > p_c$. Interestingly, the appearance of the antiferromagnetic phase in UTe$_2$ is associated with a sudden rise of the ordering temperature $T_N$ to more than 3 K, indicating a first-order character at low-temperature of the transition induced under pressure [Figure 1(a)]. A low-temperature first-order character of the transition was also reported in Ce$_{1-x}$La$_x$Ru$_2$Si$_2$ [46], suggesting that the growth of critical magnetic fluctuations may be present, at least within certain conditions, in the vicinity of first-order magnetic quantum phase transitions. This may lead to deviations from standard models initially developed for second-order quantum phase transitions [44,45].

***Magnetic and electronic properties.*** Beyond a phenomenological description of quantum magnetic criticality, a careful consideration of the electronic properties, including the magnetic properties, may be needed to describe the onset of antiferromagnetic order in UTe$_2$ under pressure. Here, we emphasize the importance of the magnetic anisotropy, magnetic exchange dimensionality, U-atoms valence, and 5$f$-electrons Fermi surface. A switch of the magnetic anisotropy was evidenced at $p_c$ by magnetic susceptibility [31] and NMR Knight-shift [48] measurements. They showed that the magnetic anisotropy, of Ising kind with the easy magnetic axis **a** for $p < p_c$, becomes of XY kind with an easy-plane ⊥ **c** for $p > p_c$. In addition, a maximum of the susceptibility $\chi_b$ measured for **H** ∥ **b** is found for $p < p_c$, while maxima of the susceptibility $\chi_a$ and $\chi_c$ measured for **H** ∥ **a**,**c**, respectively, are found for $p > p_c$. A change of dimensionality of the magnetic exchange-interaction scheme is also accompanying the onset of antiferromagnetism in UTe$_2$. A quasi-two-dimensional (2D) character of the magnetic fluctuations at ambient pressure was attributed in [25] to the absence of magnetic coupling along **c** between the U-atoms ladders. Here, signatures of a three-dimensional (3D) long-range order are indicated by resolution-limited widths in $Q_h$, $Q_k$, and $Q_l$ scans of the magnetic Bragg peaks (see Figures 2, 3, and the Supplementary Figure S2). The onset of 3D antiferromagnetic order under pressure requires the activation of a magnetic exchange along **c** between the ladders, which is presumably induced by the contraction of the lattice along **c** by pressure. A change of valence under pressure was also reported by X-ray absorption [41] and X-ray magnetic circular dichroism [49] spectroscopies. It is related with a change of the electronic delocalization, implying possible feedbacks on the magnetic properties from 5$f$-shells electrons. The $f$ electrons in heavy-fermion metals contribute to the Fermi surface, and a Fermi-surface reconstruction was observed at the magnetic quantum phase transition of several heavy-fermion antiferromagnets [50]. The Fermi surface of UTe$_2$, which was measured only at ambient pressure so far, either by angle-resolved photo-emission spectroscopy [51,52], or by de-Haas-van-Alphen experiments [53,54,55], may also be modified at the onset of antiferromagnetism under pressure. A high-pressure study of the Fermi surface may be helpful to develop band calculations describing the high-pressure antiferromagnetic phase.

***Superconductivity.*** In the first reports of superconductivity [4,5], similarities between UTe$_2$ and the U-based ferromagnetic superconductors URhGe, UCoGe, and UGe$_2$ [56] were highlighted. UTe$_2$ was presented as a nearly-ferromagnet and ferromagnetic fluctuations were proposed to drive to triplet superconductivity [5]. Here, we have found that UTe$_2$ becomes antiferromagnetic under pressure. It is a nearly-antiferromagnet at ambient pressure, and its antiferromagnetic fluctuations with wavevector **k**$_1$ become gapped in the superconducting phase SC1 [29,30]. Theoretically, antiferromagnetic fluctuations were proposed to drive to triplet superconductivity, with possible applications to UTe$_2$ and UPt$_3$, in [58], and a competition between ferromagnetic and antiferromagnetic fluctuations and their relation with superconductivity in UTe$_2$ were considered in [57]. A ferromagnetic coupling can also play a role in a nearly-antiferromagnet. In UTe$_2$, ferromagnetic coupling between U atoms within the ladders and antiferromagnetic coupling between U atoms from different ladders were proposed to drive to quasi-two-dimensional antiferromagnetic fluctuations [25]. We note that several theoretical descriptions of triplet superconductivity considered the dominant role of a ferromagnetic interaction within the U-ladder rungs [27,59].

A further challenge is to understand the relationship between the magnetic properties and the different superconducting phases induced in UTe$_2$ under pressure and magnetic field. Critical magnetic fluctuations are evidenced by the enhancement of the Fermi-liquid coefficients $\gamma$ and $A$ and of the NMR relaxation rates at $p_c$ and $H_m$ for different field directions, at ambient pressure or under pressure [6,11,16,34,60,61]. These critical magnetic fluctuations presumably drive the superconducting phases near $p_c$ and $H_m$ in UTe$_2$, but their microscopic nature is unknown. An interplay between magnetic fluctuations and superconductivity can be emphasized in other correlated-electron materials. For instance, an enhancement of antiferromagnetic fluctuations was related to that of the heat-capacity Sommerfeld coefficient $\gamma$ near the optimum doping of the high-temperature superconductor $La_{2-x}Sr_xCuO_4$ [62]. Within the conventional description introduced earlier, a possible speculation is that critical magnetic fluctuations with the wavevector $\mathbf{k_m}$ may play a role for the stabilization of the superconducting phase SC2 in $UTe_2$ near $p_c$. Under a magnetic field $\mathbf{H} \parallel \mathbf{b}$, the metamagnetic field $H_m$ collapses at $p_c$. The continuity of transition lines in the low-temperature pressure-magnetic-field phase diagram [15], but also similar NMR Knight-shift variations [18], suggests that the superconducting phase induced near to $H_m$ may be the same than the phase SC2 stabilized under pressure. Therefore, if they could be critical at $p_c$, one may also expect that magnetic fluctuations with the wavevector $\mathbf{k_m}$ could be critical at $H_m$ too. We note that magnetic-field-induced critical fluctuations, of antiferromagnetic and ferromagnetic nature, were identified at $H_m$ in $Sr_3Ru_2O_7$ [63] and $CeRu_2Si_2$ [64], respectively. Strikingly, the phase SC2 of $UTe_2$ suddenly disappears, in two very similar manners, when long-range AF order is established at $p_c$ and when a PPM regime is established at $H_m$ for $\mathbf{H} \parallel \mathbf{b}$ (see comparison of Figure 1(a) here and Figure 3(a) in [9]). Why is the phase SC2 destabilized in the AF phase under pressure and in the PPM regime for $\mathbf{H} \parallel \mathbf{b}$? And, oppositely, why is the phase SC-PPM stabilized only in the PPM regime beyond $\mu_0 H_m$ = 40 - 45 T when the magnetic field is tilted by 30 ° from $\mathbf{b}$ towards $\mathbf{c}$ [8,9]? Understanding the different domains of stability of superconductivity in $UTe_2$ and their relation with the magnetic properties offer a tough and challenging mystery to elucidate for the coming years.

**Methods**

**Sample.** The single crystal of $UTe_2$, of dimensions 2.7*1.7*1.3 mm$^3$ and mass of 54 mg, studied here was grown by chemical vapor transport as reported in [5]. Iodine was used as transport agent with a ratio of 3 mg/cm$^3$ relative to quartz ampoule volume. Uranium to Tellurium ratio ranged from 1.65 to 1.9. The ampoule was carefully outgazed prior sealing and closed under secondary vacuum. The sample had a superconducting transition temperature $T_{sc}$ = 1.7 K.

**Pressure cell.** High-pressure neutron-diffraction experiments were performed in a hybrid CuBe/NiCrAl clamp-piston-cylinder cell. As pressure medium we used deuterated glycerin U-D8 (99% Atom D). Glycerin generally provides very good hydrostaticity up to 5 GPa. The pressure was monitored by the superconducting transition of a lead sample, which was measured by ac susceptibility. The pressure was fixed prior the first neutron diffraction experiment on D23, and checked again before the experiment on WISH. The accuracy of the pressure determination is ± 0.03 GPa and no pressure change was observed between the different experiments. The pressure cell was slowly cooled to low temperatures, to avoid stress on the sample due to the solidification of the glycerin. The pressure was further confirmed by the values of the lattice parameters obtained by neutron diffraction (see Supplementary Figure S1).

**Neutron diffraction.** Most of the data presented here were collected using WISH, which is a long-wavelength time-of-flight neutron-diffraction spectrometer on the second-target station at the spallation source ISIS, Didcot. Its pixelated detector coverage, of ≈ 340 ° in the plane and ±15 ° out of plane allows a large portion of reciprocal space to be obtained in a single crystal orientation. Measurements were done with a neutron wavelength $\lambda$ in the range 0.7 - 10 Å and the basal plane perpendicular to the axis $\mathbf{a}$ of the crystal. Two configurations, with the incident neutron beam making an angle of 14.5 and 38.5 °, respectively, with the axis $\mathbf{c}$ of the crystal, were investigated. Data analysis was performed with MANTID. Complementary measurements were made using the double-axis diffractometer D23 at the ILL, Grenoble. A configuration with a neutron wavelength $\lambda$ = 1.283 Å from a copper monochromator was used and the studied crystal was mounted in a cryostat with the basal plane perpendicular to the axis **a**. Both series of neutron experiments were performed on the same sample at the pressure $p$ = 1.8 GPa, in the same pressure cell.

Supplementary References concerning the Section 'Method' are given in the Supplementary Information.

## Data availability

The data that support the findings of this study are available from the corresponding author on reasonable request. Raw data collected on D23 at the ILL can be found at https://doi.ill.fr/10.5291/ILL-DATA.CRG-2971.

## Acknowledgments

We acknowledge useful discussions with Jean-Pascal Brison, Jacques Flouquet and Charles Simon. We received financial support from the French Research Agency ANR within the projects FRESCO No. ANR-20-CE30-0020, from the Fédération Française de Diffusion Neutronique (2FDN), from the JPSJ programs KAKENHI P22H04933, JP20K20889, JP20H00130, JP20KK0061, and from the "Programme Investissements d'Avenir" under the project ANR-11-IDEX-0002-02 (reference ANR-10-LABX-0037-NEXT).

## Author contributions

The studied sample was grown by G.L. in close collaboration with D.A. It was characterized at ambient pressure by G.L., D.B. and G.K. The pressure cell was prepared by G.K. and D.B. Neutron diffraction experiments using WISH at ISIS were performed by W.K., T.T., S.R, P.M., D.D.K., and F.O. and neutron diffraction experiments using D23 at the ILL were performed by W.K., T.T., S.R, G. K., E.R. and K.B. Data were analyzed by W.K, T.T., P.M., D.D.K., F.O., E.R., and S.R. The paper was written by W.K. with contributions from all of the authors.

**Competing interests:** The authors declare no competing interests.

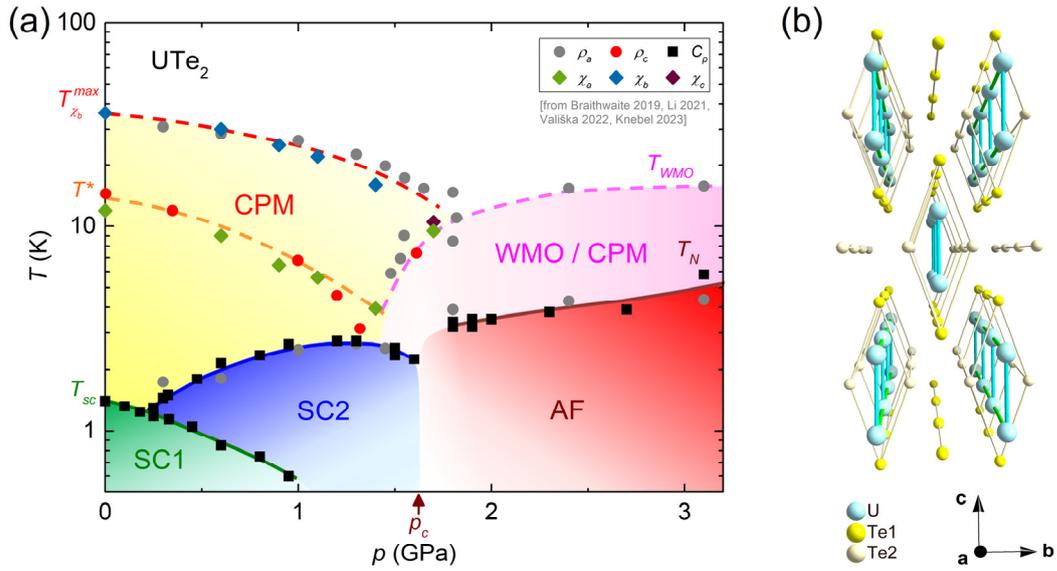

**Figure 1: Pressure-temperature phase diagram and crystal structure of UTe$_2$.** (a) Pressure-temperature phase diagram constructed from electrical-resistivity $\rho_a$ and $\rho_c$ measurements with currents **I** ∥ **a**,**c**, respectively (from [32,34]), heat-capacity $C_p$ measurements (from [6]), and magnetic susceptibility $\chi_a$, $\chi_b$ and $\chi_c$ measurements in magnetic fields **H** ∥ **a**,**b**,**c**, respectively (from [31]). Pressures have been slightly rescaled to combine the different sets of data. CPM denotes the correlated paramagnetic regime, WMO/CPM the high-pressure weak-magnetic-order / correlated paramagnetic regime, SC1 and SC2 the low-pressure and pressure-induced superconducting phases, and AF the antiferromagnetic phase studied here. (b) Crystal structure extended over several unit cells emphasizing the two-legs ladder arrangement of U atoms. The shortest U-U distance corresponds to the rungs of the ladders.

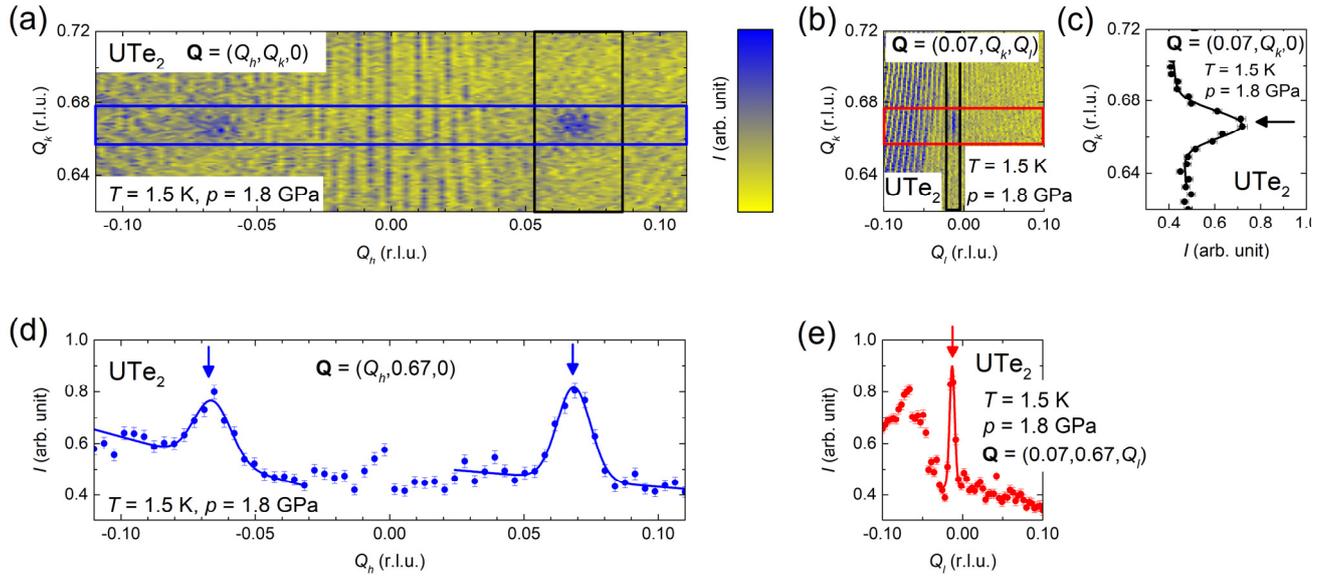

**Figure 2: Magnetic Bragg peaks at the momentum transfers Q = (±0.07,0.67,0) at $T$ = 1.5 K.** Neutron scattered intensity maps (a) in the ($Q_h$,$Q_k$,0) plane and (b) in the (0.07,$Q_k$,$Q_l$) plane. (c) $Q_k$, (d) $Q_h$, and (e) $Q_l$ scans extracted after integration with windows of widths $\Delta Q_h$ = 0.033, $\Delta Q_k$ = 0.02, and $\Delta Q_l$ = 0.017 along the $Q_h$, $Q_k$, and $Q_l$ directions, respectively. The lines show fits to the data by a Gaussian function with a linear background. The windows of integration are indicated by blue, black, and red lines in the maps shown in (a-b).

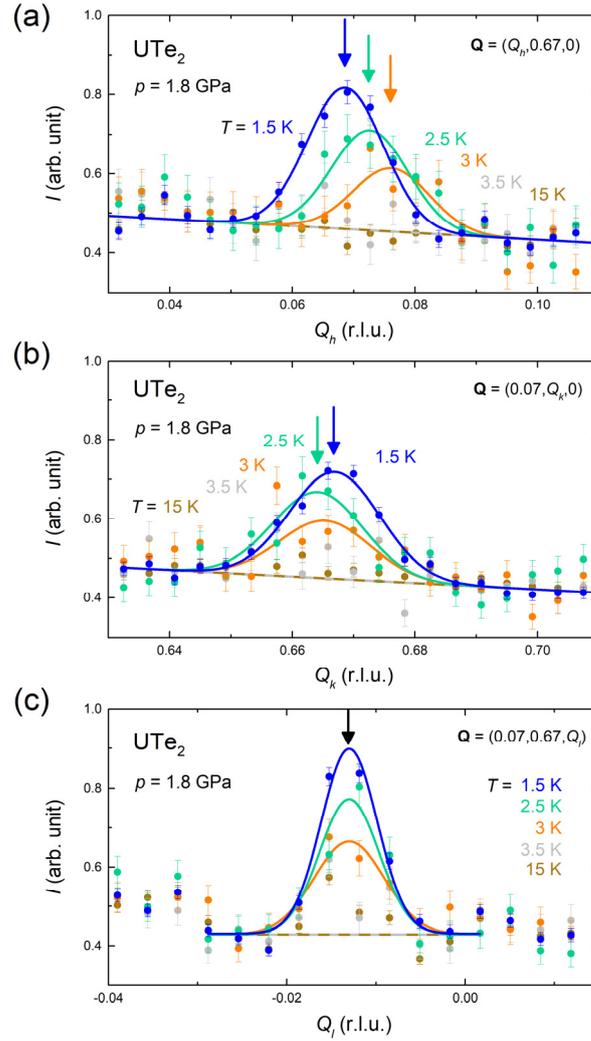

**Figure 3: Temperature variation of the magnetic Bragg peak at the momentum transfer Q = (0.07,0.67,0).** (a) $Q_h$, (b) $Q_k$, and (c) $Q_l$ scans extracted with windows of widths $\Delta Q_h$ = 0.033, $\Delta Q_k$ = 0.02, and $\Delta Q_l$ = 0.017, respectively [the windows of integration are indicated by blue, black, and red lines in the maps shown in Figures 1(a-b)], at temperatures 1.5 K $\leq T \leq$ 15 K. The lines show fits to the data by a Gaussian function with a linear background.

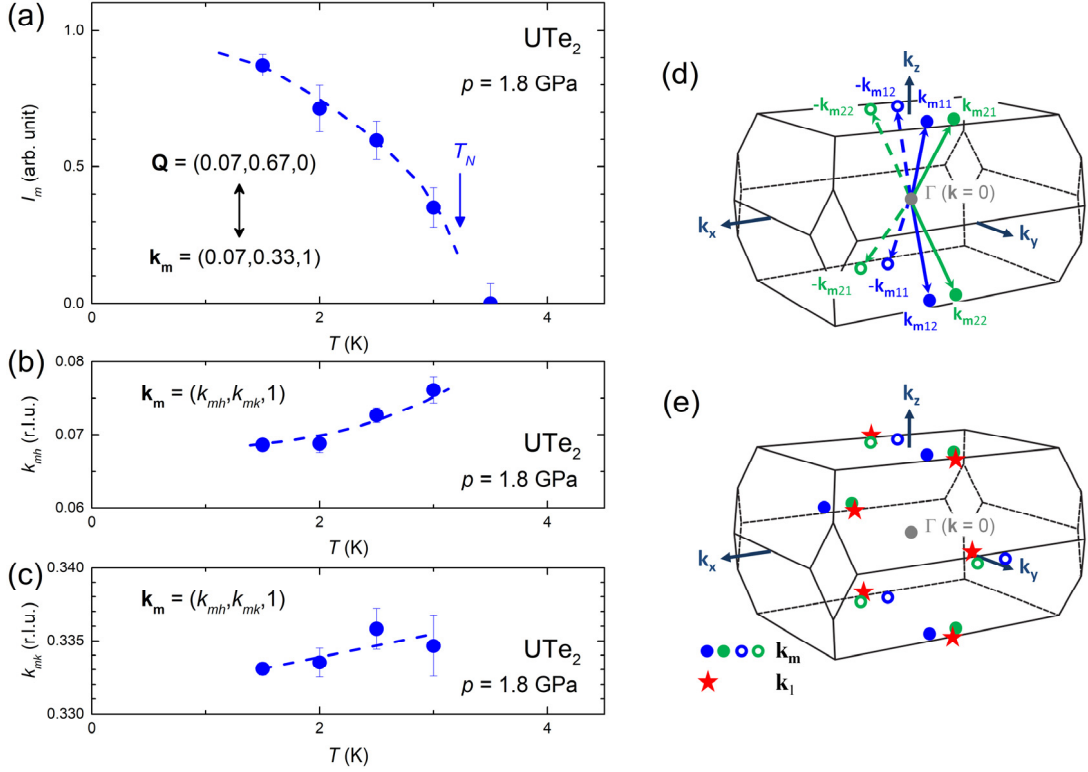

**Figure 4: Temperature dependence of intensity and momentum-transfer components of the magnetic Bragg peak and its equivalent positions in the FBZ.** Temperature dependence of (a) the intensity of the magnetic Bragg peak deduced from integration of the peak from *d*-spacing scans (see Supplementary Information and Supplementary Figure S3), and the (b) $k_{mh}$ and (c) $k_{mk}$ components of the extracted wavevector $\mathbf{k_m} = (k_{mh}, k_{mk}, 1)$. (d) Position of the equivalent magnetic wavevectors $\mathbf{k}_{mij} = (\pm 0.07, \pm 0.33, \pm 1)$ on the border of the FBZ of UTe2. (e) Comparison, within and near the FZB, of the positions of equivalent wavevectors $\mathbf{k_m}$ characterizing the antiferromagnetic phase under pressure and $\mathbf{k_1}$ at which antiferromagnetic fluctuations were reported at ambient pressure.

# Supplementary Information

# Incommensurate antiferromagnetism in UTe$_2$ under pressure

W. Knafo,[1,*] T. Thebault[1], P. Manuel[2], D.D. Khalyavin[2], F. Orlandi[2], E. Ressouche[3],
K. Beauvois[3], G. Lapertot[4], K. Kaneko[5,6], D. Aoki[7], D. Braithwaite[4], G. Knebel[4], S. Raymond[3]

[1] *Laboratoire National des Champs Magnétiques Intenses - EMFL, CNRS, Université Grenoble Alpes, INSA-T, Université Toulouse 3, 31400 Toulouse, France.*
[2] *ISIS Pulsed Neutron and Muon Source, STFC Rutherford Appleton Laboratory, Harwell Science and Innovation Campus, Oxfordshire OX11 0QX, United Kingdom.*
[3] *Université Grenoble Alpes, CEA, IRIG, MEM, MDN, 38000 Grenoble, France.*
[4] *Université Grenoble Alpes, CEA, Grenoble INP, IRIG, PHELIQS, 38000 Grenoble, France.*
[5]*Materials Sciences Research Center, Japan Atomic Energy Agency, Tokai, Ibaraki 319-1195, Japan*
[6]*Advanced Science Research Center, Japan Atomic Energy Agency, Tokai, Ibaraki 319-1195, Japan*
[7] *Institute for Materials Research, Tohoku University, Ibaraki 311-1313, Japan*

\* *Corresponding author: william.knafo@lncmi.cnrs.fr*

**Supplementary Discussion**

In the main text, we mention that, due to the limited number of magnetic Bragg peaks extracted here, we cannot determine the magnetic structure. We cannot conclude whether the magnetic structure of UTe$_2$ under pressure is of spin-density-wave kind, i.e., with magnetic moments varying as:

$$\boldsymbol{\mu}(\mathbf{R}_{ij}) = \mu_u \cos(2\pi \mathbf{k_m} \mathbf{R}_i + \phi) \mathbf{u} \qquad (S1)$$

or of helical kind, i.e., with magnetic moments varying as:

$$\boldsymbol{\mu}(\mathbf{R}_{ij}) = \mu_u \cos(2\pi \mathbf{k_m} \mathbf{R}_i + \phi) \mathbf{u} + \mu_v \sin(2\pi \mathbf{k_m} \mathbf{R}_i + \phi) \mathbf{v} \qquad (S2)$$

Here, $j$ labels one of the two U atoms, which have in-phase magnetic moments, in the primitive cell $i$. $\mathbf{R}_{ij} = \mathbf{R}_i + \mathbf{r}_j$ is the position of the atom $j$, $\mathbf{R}_i$ is the position at the center of the primitive cell $i$, $\mathbf{r}_j$ is the relative position of the atom $j$ in the primitive cell, and $\mathbf{u}$ and $\mathbf{v}$ are two orthogonal directions. A spin-density-wave structure can be seen as the limiting case of a helical structure for $\mu_v \rightarrow 0$. For instance, a spin-density wave structure was found in URu$_2$Si$_2$ in its ordered phase stabilized in high magnetic fields from 35 to 39 T applied along **c** [65], and a helical structure was found in CeRhIn$_5$ [66].

The amplitude $\mu_\perp$ = 0.3 ± 0.05 $\mu_B$ / U estimated from the Bragg peaks intensity measured here corresponds to the projection of the antiferromagnetic moment $\mu = (\mu_u^2 + \mu_v^2)^{1/2}$ perpendicular to the investigated momentum transfers **Q**, i.e., mainly perpendicular to the direction **b**.

We also concluded, from the observation of magnetic Bragg peaks at momentum transfers of components $Q_l = 0$, that the two closest U atoms, which form the rungs of the ladders, have in-phase magnetic moments. This information directly follows from the magnetic structure factors of these reflections, since only the coordinates $z$ of the two U atoms - which are distant by $d_1$ - in a primitive cell are different (see [25]). The neutron diffracted intensity is proportional to the square $F_m^2$ of the magnetic structure factor, and $F_m$ is proportional to $\cos(\pi Q_l d_1/c)$, which is equal to 1 for $Q_l = 0$, in the case of an in-phase arrangement of the two closest U atoms. Oppositely, $F_m$ would be proportional to $\sin(\pi Q_l d_1/c)$, which is equal to 0 for $Q_l = 0$, in the case of an out-of-phase arrangement of the two closest U atoms. This last case can be excluded since diffracted intensities were measured at $Q_l = 0$.

**Supplementary Methods**

**Pressure cell.** Supplementary Reference [67] shows a similar cell than that used in the present study. It is demonstrated in Supplementary Reference [68] that glycerin generally provides very good hydrostaticity up to 5 GPa.

**Neutron diffraction.** Supplementary Reference [69] presents details about the long-wavelength time-of-flight neutron-diffraction spectrometer WISH, and Supplementary Reference [70] presents the program MANTID used to analyze the data collected on WISH.

**Supplementary Figures**

Supplementary Figure S1 shows that the lattice parameters determined in the present neutron diffraction experiments are consistent with the estimated pressure $p = 1.8$ GPa.

Supplementary Figure S2 shows data collected at the momentum transfers **Q** = (±0.07,1.33,0) and at the temperature $T = 1.5$ K, within the second experimental configuration, with an angle of 38.5 ° between the incident neutron beam and **c**, on the time-of-flight spectrometer WISH.

Supplementary Figure S3 shows $d$-spacing scans on the magnetic Bragg peaks at **Q** = (0.07,0.67,0), whose integration was used to extract the Bragg peak intensity $I_m$, from data collected over a selection of detectors pixels at different temperatures from 1.5 to 3.5 K. The selection of pixels was adjusted to cover the small variation with temperature of the magnetic wavevector [see Figures 4(b,c)]. $d$-spacing scans result from diffracted neutrons with different wavelengths, i.e., with different times of flight to the detectors ($d$ is the distance defined by $\lambda = 2d\sin\theta$, where $\lambda$ is the neutron wavelength and $\theta$ is the scattering angle).

Supplementary Figures S4 and S5 show scans on magnetic Bragg peaks and structural Bragg peaks, respectively, measured complementarily using the two-axis diffractometer D23.

Supplementary Figure S6 shows a comparison of the temperature dependences of the magnetic Bragg peak intensity, extracted from the integration of $d$-spacing scans from the experiment using WISH, or from the intensity directly measured at the Bragg position from the experiment using D23.

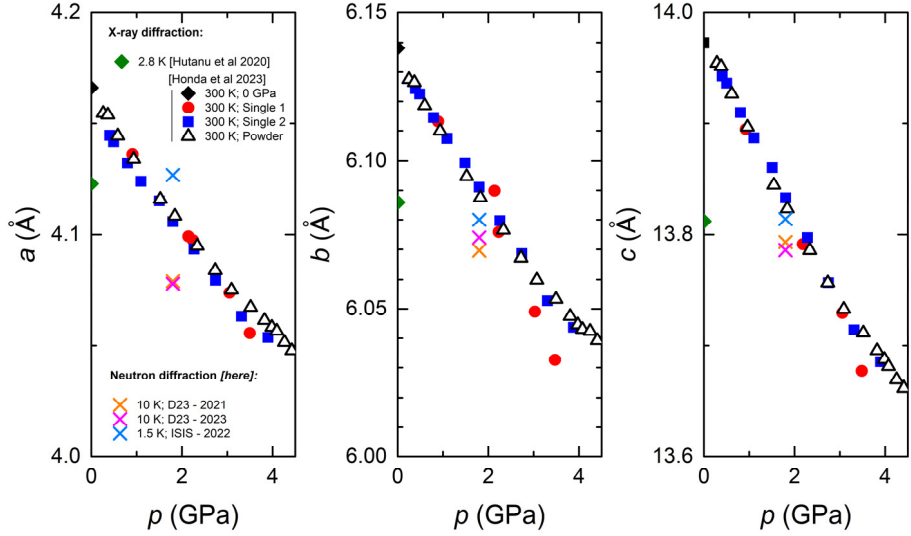

**Supplementary Figure S1: Lattice parameters a, b, and c of UTe$_2$.** Lattice parameters extracted by x-ray diffraction at $T = 2.8$ K and $p = 1$ bar in [71], at $T = 300$ K under pressure up to 4.5 GPa in [14], and extracted by neutron diffraction here, from three experiments at $T = 10$ K on the spectrometer D23 at the ILL and at $T = 1.5$ K at the spectrometer WISH at ISIS, performed with the same sample in the same pressure cell and at the same pressure $p = 1.8$ GPa.

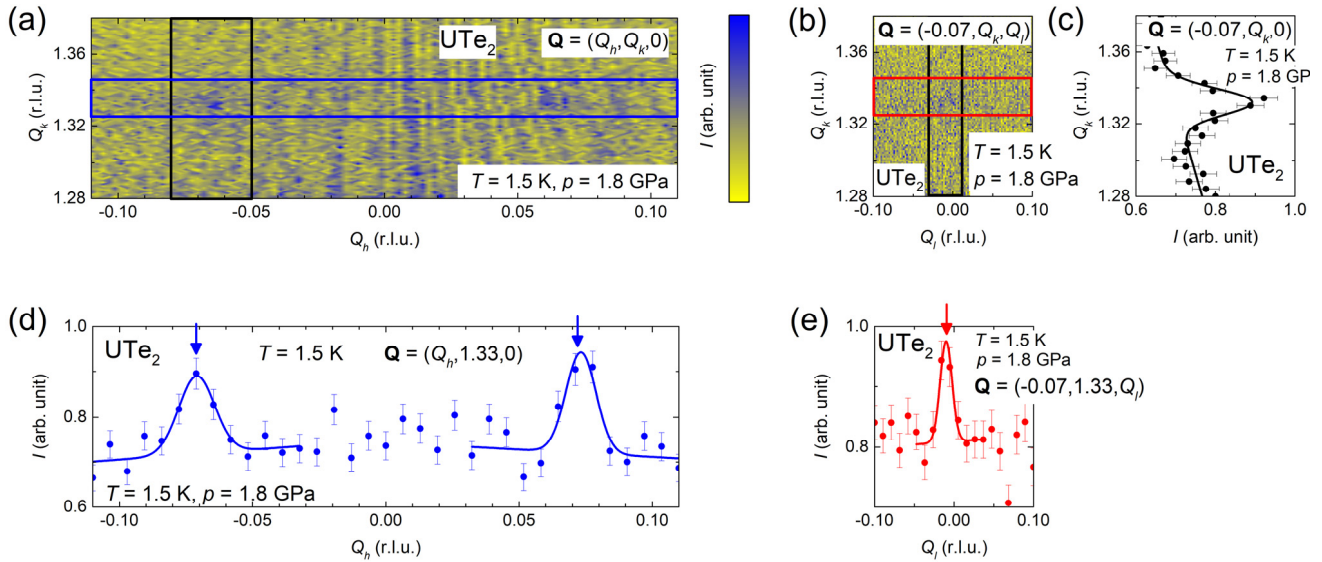

**Supplementary Figure S2: Magnetic Bragg peaks at the momentum transfers Q = ($\pm$0.07,1.33,0) at $T = 1.5$ K.** Neutron scattered intensity maps (a) in the $(Q_h,Q_k,0)$ plane and (b) in the $(0.07,Q_k,Q_l)$ plane. (c) $Q_k$, (d) $Q_h$, and (e) $Q_l$ scans extracted after integration with windows of widths $\Delta Q_h = 0.03$, $\Delta Q_k = 0.02$, and $\Delta Q_l = 0.04$ along the $Q_h$, $Q_k$, and $Q_l$ directions, respectively. The lines show fits to the data by a Gaussian function with a linear background. The windows of integration are indicated by blue, black, and red lines in the maps shown in (a-b).

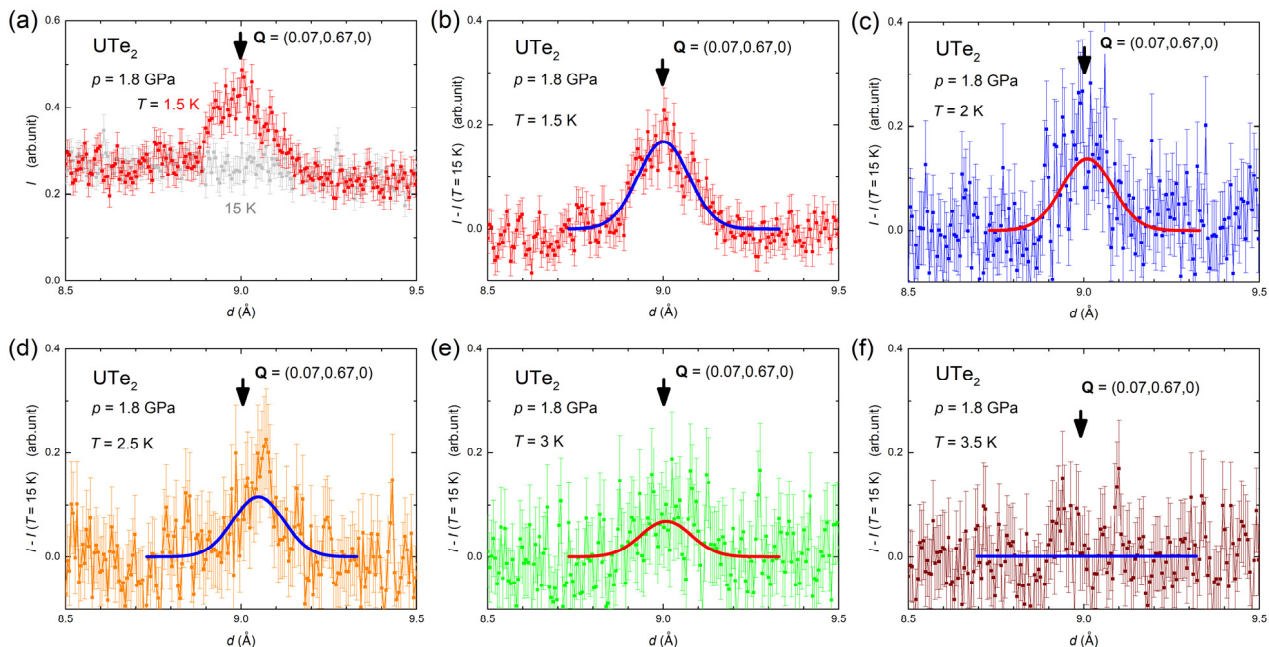

**Supplementary Figure S3: Temperature-dependence of *d*-spacing scans on the magnetic Bragg peaks at Q = (0.07,0.67,0)**. *d*-spacing scans at the temperatures (a) $T$ = 1.5 and 15 K, (b) $T$ = 1.5 K, (c) $T$ = 2 K, (d) $T$ = 2.5 K, (e) $T$ = 3 K, and (f) $T$ = 3.2 K. The lines show fits to the data by a Gaussian function.

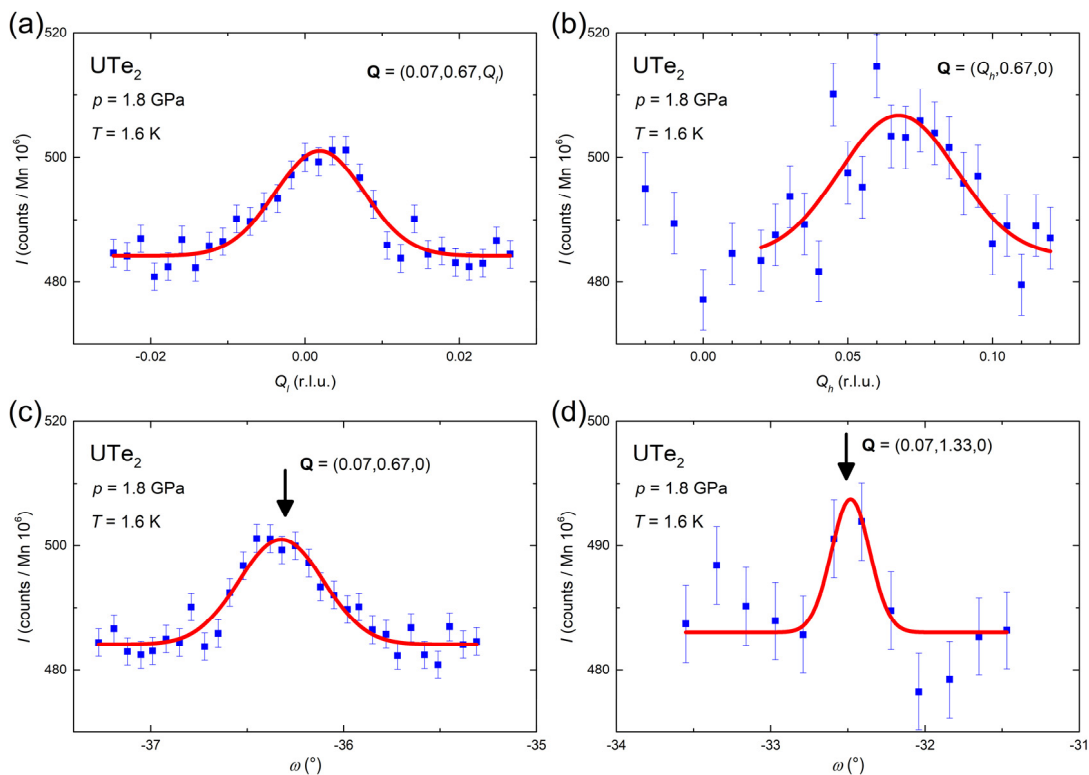

**Supplementary Figure S4: Magnetic Bragg peaks measured on D23 at $T$ = 1.6 K**. (a) $Q_l$, (b) $Q_h$, and (c) $\omega$ scans measured at Q = (0.07,0.67,0) and (d) $\omega$ scan measured at Q = (0.07,1.33,0).

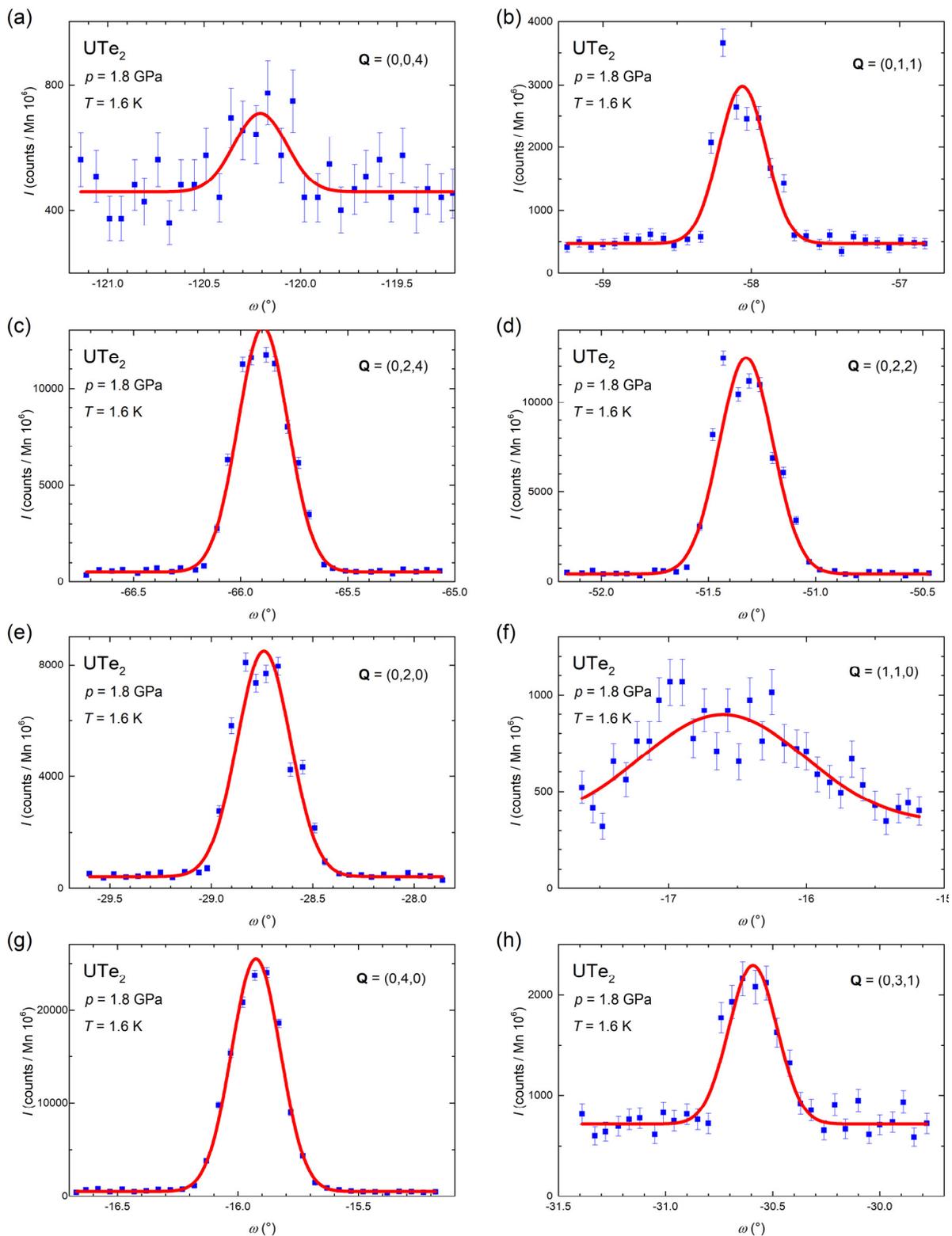

**Supplementary Figure S5: Structural Bragg peaks measured on D23 at $T$ = 1.6 K.** $\omega$ scans measured at the momentum transfers (a) $\mathbf{Q}$ = (0,0,4), (b) $\mathbf{Q}$ = (0,1,1), (c) $\mathbf{Q}$ = (0,2,4), (d) $\mathbf{Q}$ = (0,2,2), (e) $\mathbf{Q}$ = (0,2,0), (f) $\mathbf{Q}$ = (1,1,0), (g) $\mathbf{Q}$ = (0,4,0), (h) $\mathbf{Q}$ = (0,3,1).

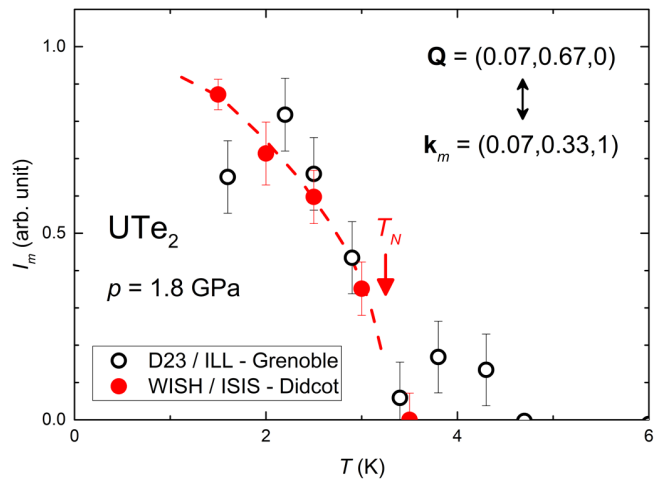

**Supplementary Figure S6: Temperature dependence of the intensity of the magnetic Bragg peak extracted at the momentum transfer Q = (0.07, 0.67, 0) at the diffractometers D23 and WISH.** The intensity was deduced from an integration of the peak in *d*-spacing scans (see Supplementary Figure S3) for the experiment made at WISH, and from a direct counting at the Bragg position at D23.

**Supplementary References**